\documentclass{ws-p8-50x6-00}

\newcommand{\beq}{\begin{eqnarray}}
\newcommand{\eeq}{\end{eqnarray}}

\newcommand{\ie}{{\it i.e.\ }}

\newcommand{\aphi}{\langle \phi \rangle}
\newcommand{\pbp}{\langle \bar{\psi} \psi \rangle}

\begin{document}

\title{Imaginary chemical potential in QCD at finite temperature}

\author{Massimo D'Elia}

\address{Dipartimento di Fisica dell'Universit{\`a} di
    Genova and INFN, I-16146, Genova, Italy\\ 
E-mail: delia@ge.infn.it}

\author{Maria-Paola Lombardo}

\address{Istituto Nazionale di Fisica Nucleare, Sezione di Padova, 
I-35131, Padova, Italy\\
E-mail: lombardo@pd.infn.it}  

\maketitle

\abstracts{
After presenting a brief review of how simulations of QCD with
imaginary chemical potential can be used to extract physical results,
we analyse the phase structure of QCD
with four flavours of dynamical fermions in the finite temperature - imaginary
chemical potential plane, and discuss perspectives for realistic calculations.}

\section{Introduction}

The zero density QCD partition function,
$Z(V,T) = {\rm Tr} \left( e^{-\frac{H_{\rm QCD}}{T}} \right) $,
with $H_{\rm QCD}$ the QCD Hamiltonian,
can be discretized on an euclidean lattice with finite temporal extent
$\tau = 1/T$
\beq
Z = \int  \left( {\cal D}U {\cal D}\psi {\cal D}\bar{\psi} \right)
e^{-\beta S_G[U]}e^{-S_F[U,\psi,\bar{\psi}]} = 
\int  \left( {\cal D}U \right) e^{-\beta S_G[U]} \det M[U]
\label{lattpf}
\eeq
where $U$ are the gauge link variables and $\psi$ and $\bar{\psi}$ the fermionic variables, 
$S_G$ is the pure gauge action, $S_F$ is the fermionic action which can be expressed
as a quadratic form in the fermionic fields in terms of the fermionic matrix
$M[U]$, $S_F = \bar{\psi} M[U] \psi$.

To describe QCD at finite density the grand canonical partition function,
$Z(V,T,\mu) = {\rm Tr} \left( e^{-\frac{H_{\rm QCD} - \mu N}{T}} \right) $,
where $N = \int d^3 x \psi^\dagger \psi$ is the quark number operator, can be used. 
The correct way to introduce
a finite chemical potential $\mu$ on the lattice~\cite{kar} is to modify 
the temporal links appearing in the integrand in Eq.~(\ref{lattpf}) as follows:
\beq
U_t &\to& e^{a \mu} U_t \;\;\;\;\;\;\;\; {\rm (forward\; temporal\; link)} \nonumber \\
U_t^\dagger &\to& e^{- a \mu} U_t^\dagger \;\;\;\; {\rm (backward\; temporal\; link)}\; ,
\label{linktr}
\eeq
where $a$ is the lattice spacing.
Whilst $S_G$ is left invariant by this transformation, $\det M[U]$ is not and 
gets  a complex phase which makes importance sampling, and therefore
standard lattice MonteCarlo simulations, unfeasible.

The situation is different when the chemical potential is purely imaginary.
This is implemented on the lattice as described in Eq.~(\ref{linktr}),
but in this case $U_t \to e^{i a \mu_I} U_t$, 
$U_t^\dagger \to e^{- i a \mu_I} U_t^\dagger$.
This is like adding a constant $U(1)$ background field to the original theory; 
$\det M[U]$ is again real and positive and simulations are as easy
as at $\mu = 0$.

The question then arises how simulations at imaginary chemical potential may be
of any help to get physical interesting information.

One possibility is analytic continuation~\cite{mpl}.
$Z(V,T,\mu)$ is expected to be an analytical even function of $\mu$
away from phase transitions. For small enough $\mu$ one can write:
\beq
\log Z(\mu) &=& a_0 + a_2 \mu^2 + a_4 \mu^4 + O(\mu^6) \; \\
\log Z(\mu_I) &=& a_0 - a_2 \mu_I^2 + a_4 \mu_I^4 + O(\mu_I^6) \; .
\label{taylor}
\eeq
Simulations at small $\mu_I$ will thus allow a determination of the expansion 
coefficients for the free energy and, analogously, for other physical quantities, which can be cross-checked with those obtained by standard reweighting
techniques~\cite{glasgow,foka}.
This method is expected to be useful in the  high temperature regime, where 
the first coefficients should be sensibly different from zero; moreover
the region of interest for present experiments (RHIC, LHC) is that of 
high temperatures and small chemical potential, with $\mu / T \sim 0.1$.
This method has been already investigated in the strong coupling regime~\cite{mpl}, in the dimensionally reduced $3$--$d$ QCD theory~\cite{hart1},
and in full QCD with two flavours~\cite{deph}.
The Taylor expansion coefficients can also be measured as
derivatives with respect to $\mu$ at $\mu = 0$ ~\cite{susce,deriv}.

$Z(V,T,i \mu_I)$ can also be used to reconstruct
the canonical partition function $Z(V,T,n)$ at fixed quark number $n$~\cite{roberge},
\ie at fixed density:
\beq
Z(V,T,n) &=& {\rm Tr} \left( ( e^{-\frac{H_{\rm QCD}}{T}} \delta(N - n) \right) = 
\frac{1}{2\pi} {\rm Tr} \left( e^{-\frac{H_{\rm QCD}}{T}} \int_0^{2\pi} 
{\rm d} \theta e^{i \theta (N - n)} \right) \nonumber \\
&=& \frac{1}{2\pi} \int_0^{2\pi}
{\rm d} \theta e^{- i \theta n} Z(V,T,i \theta T) \; .
\label{canonical}
\eeq
As $n$ grows, the factor
$e^{- i \theta n}$ oscillates more and more rapidly and the error in the 
numerical integration grows exponentially with $n$: this makes the application
of the method difficult especially at low temperatures where $Z(V,T,i \mu_I)$ 
depends very weakly on $\mu_I$.
The method has been applied in the 2--d Hubbard model~\cite{alford},
where $Z(V,T,n)$ has been reconstructed up to $n = 6$.

The phase structure of QCD in the $T$ -- $i \mu_I$ plane is also interesting
by its own. Writing for brief $Z(\theta) \equiv Z(V,T,i \theta T) = 
{\rm Tr} \left( e^{ i \theta N} e^{-\frac{H_{\rm QCD}}{T}} \right) $, $Z(\theta)$ 
is clearly periodic in $\theta$ with period $2 \pi$ and a period 
$2 \pi /3$ is expected in the confined phase, where 
only physical states with $N$ multiple of 3 are present.
However it has been shown~\cite{roberge} that $Z(\theta)$ has always period 
$2 \pi /3$ for any physical temperature.
Moreover the suggestion has been made~\cite{roberge}, based on a calculation
in the weak  coupling approximation, that  discontinuities 
in the first derivatives of the free energy at $\theta = 2 \pi /3 (k + 1/2)$
should appear in the high temperature phase.
This suggests a very interesting scenario for the phase diagram of QCD in 
the $T$ -- $i \mu_I$ plane which needs confirmation by lattice calculations.

We have recently started a program of simulations of QCD at finite imaginary
chemical potential, with both the aim of 
determining the phase diagram of QCD in the entire
$T$ -- $i \mu_I$ plane, and exploring by analytic continuation the
high T -- small real chemical potential region.

We have studied QCD with four degenerate staggered flavours of
bare mass $a \cdot m = 0.05$ on a $16^4 \times 4$ lattice, where the  phase
transition is expected at a critical coupling 
$\beta_c \simeq 5.04$~\cite{brown} 
(the two flavour case
has been studied as well~\cite{deph}). The algorithm used is the standard HMC $\Phi$ algorithm.
We will present here only a subset of our results and analysis.
A complete presentation will appear soon~\cite{else}.

\section{Results}

In order to understand the phase structure of the theory, it is very useful
to look at the phase of the trace of the Polyakov loop, $P(\vec{x})$. 
Let us parametrize $P(\vec{x}) \equiv |P(\vec{x})| e^{i \phi}$,
and let $\aphi$ be the average value of the phase.
In the pure gauge theory the average Polyakov loop is non zero only in the
deconfined phase, where the center symmetry is spontaneously broken
and $\aphi = 2 k \pi / 3$, $k = -1,0,1$, \ie the Polyakov
loop effective potential is flat in the confined phase and develops 
three degenerate minima above the critical temperature.
In presence of dynamical fermions $P(\vec{x})$ enters explicitely
the fermionic determinant and $Z_3$ is broken: the effect of the determinant
is therefore like that of an external magnetic field which 
aligns the Polyakov loop along $\aphi = 0$. In the high temperature
phase the $Z_3$ degeneracy is lifted: $\aphi = 0$ is the true vacuum
and  $\aphi = \pm 2 \pi /3$ are now metastable minima.

When $\mu_I \neq 0$, what enters the fermionic
determinant is $P(\vec{x}) e^{i \theta}$, $\theta \equiv \mu_I/T$, 
instead of $P(\vec{x})$.
Therefore the determinant now tends to
align $\aphi + \theta$ along zero: its effect is 
like that of an external magnetic field pointing in the 
$\theta$ direction. Hence one expects $\aphi = - \theta$ at
low temperatures; at high temperatures the external magnetic field
will still lift the $Z_3$ degeneracy, but now which is the true vacuum
will depend on the value of $\theta$. In particular one expects 
that for $ (k - 1/2) < \frac{3}{2 \pi} \theta < (k + 1/2)$ the true
vacuum is the one with $\aphi =  2 k \pi / 3$ and that 
$\theta =  2 (k + 1/2) \pi / 3$ corresponds to first order 
phase transitions from one $Z_3$ sector to the other: 
this is indeed the prediction of Roberge and Weiss.

\begin{figure}[!t]
\vspace{0cm}
\epsfxsize=25pc 
\epsfbox{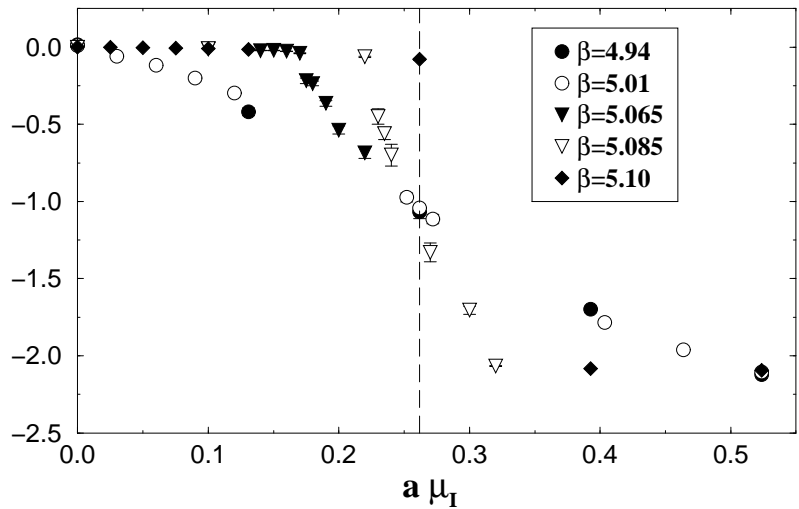}
\vspace{-0.30cm}
\caption{Average value of the Polyakov loop phase as a function
of the imaginary chemical potential for different values of $\beta$.
The vertical dashed line corresponds to $\theta = \mu_I/T = \pi/3$.}
\label{phasefig}
\vspace{-0.6cm}
\end{figure}

In Fig.~\ref{phasefig} we report our results for $\aphi$ versus the
imaginary chemical potential for different values of $\beta$. Since
$T = 1/(N_t a)$ and $N_t = 4$ in our case, we have $\theta = 4 a \mu_I$.
For $\beta = 4.94$ and $5.01$, which are below the critical $\beta$
at $\mu_I = 0$, $\beta_c(\mu_I=0) \equiv \beta_c \simeq 5.04$, one has 
$\aphi \simeq - \theta = - 4 a \mu_I$, \ie $\aphi$ is driven continously 
by the fermionic determinant. For $\beta = 5.10$, which is well
above $\beta_c$ we see that $\aphi \simeq 0$, almost independently of
$\mu_I$, as long as $\theta < \pi/3$, while for $\theta > \pi /3$
there is a sudden change to $\aphi \simeq - \pi/3$: we are clearly
crossing the Roberge-Weiss phase transition from one $Z_3$ sector
to the other. 
At intermediate values, $\beta = 5.065$ and $5.085$, $\aphi \simeq 0$
until a critical value of $a \mu_I$, where it starts moving almost 
linearly with $\mu_I$ crossing continuously the $Z_3$ boundary: in this case
there is no Roberge-Weiss phase transition, but there is anyway a critical value
of $\mu_I$ after which $\aphi$ is no more constrained to be $\simeq 0$
and can be driven again by $\theta$: as we will soon clarify,
this critical value of $\mu_I$ corresponds to the crossing of the 
chiral critical line, 
\ie the continuation in the $T$--$\mu_I$ plane of the chiral phase transition.

\begin{figure}[!t]
\vspace{0cm}
\epsfxsize=26pc 
\epsfbox{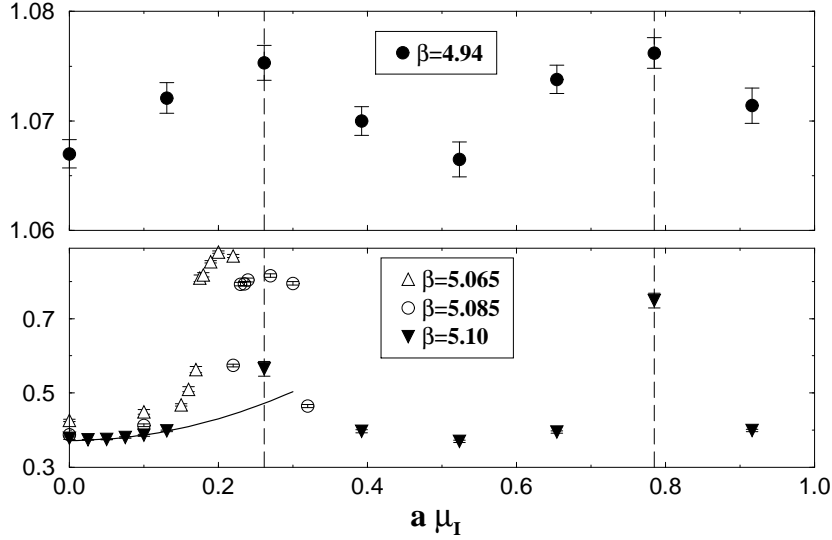}
\vspace{-0.2cm}
\caption{Average value of the chiral condensate as a function
of the imaginary chemical potential for different values of $\beta$.
The vertical dashed lines correspond to $\theta = \mu_I/T = (2 k + 1) \pi/3$.
The continuous line in the lower picture is the result of a 
quadratic fit at small values of $a \mu_I$ obtained at $\beta = 5.10$.}
\label{psifig}
\vspace{-0.6cm}
\end{figure}

\begin{figure}[!t]
\vspace{-0.60cm}
\epsfxsize=25pc 
\epsfbox{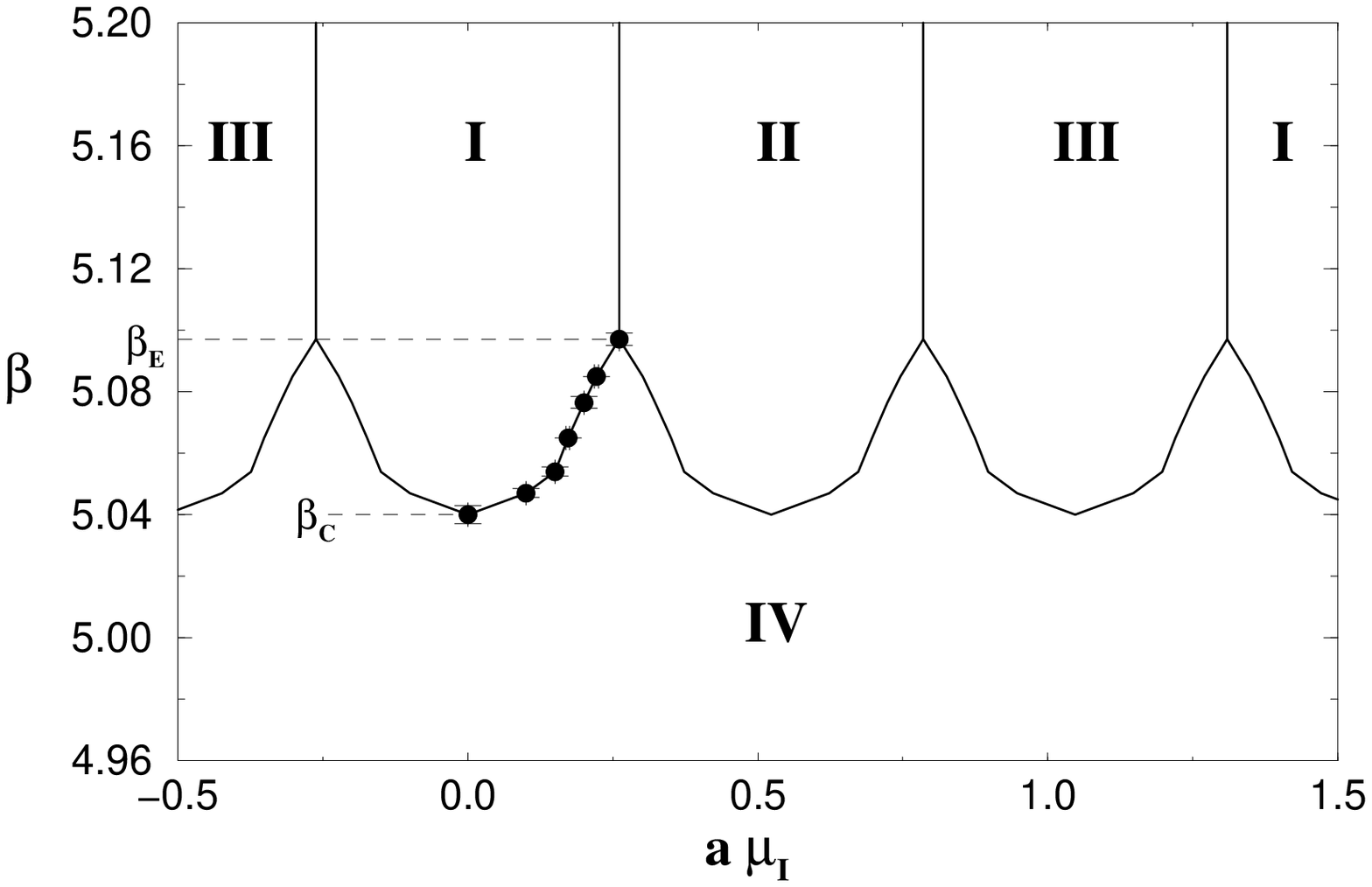}
\vspace{-0.30cm}
\caption{Sketch of the phase diagram in the $\mu_I$--$\beta$ plane.
The filled cirles represents direct determinations of the chiral critical 
line location from our simulations. The rest of the chiral line 
has been obtained by interpolation and by exploiting
the symmetries of the partition function.}
\label{diagfig}
\vspace{-0.6cm}
\end{figure}

We display our results for the chiral condensate in Fig.~\ref{psifig}. 
We expect a periodicity
with period $2\pi /3$ in terms of $\theta$. Moreover $\pbp$, like
the partition function, is an even function of $\mu_I$: 
this, combined with the periodicity, leads to
symmetry around all points $\theta = n \pi/3$.
with $n$ an integer number, for $\pbp$ as well as for the partition
function itself.
For $\beta < \beta_c$, $\pbp$ has a continuous dependence on 
$a \mu_I$ with the expected periodicity and symmetries.
For $\beta > \beta_c$ the correct periodicity and symmetries are
still observed but the dependence is less trivial. 
At $\beta = 5.065$ there is a critical value $a \mu_I \simeq 0.17$ for which 
the theory has a transition to a spontaneously broken chiral
symmetry phase: we are clearly going through the chiral critical line. 
The same happens for $\beta = 5.085$ at $a \mu_I \simeq 0.22$: in this 
case we have proceeded further, observing also the transition back
to a chirally restored phase at $a \mu_I \simeq 0.30$, which is, correctly,  
the symmetric point with respect to $\theta = \pi / 3$\footnote{
We notice a discrepancy in the results obtained at the same
$\beta$ by multireweighting techniques~\cite{foka},
where the same symmetry cannot be observed.}.
At $\beta = 5.10$ we never cross, when moving in $\mu_I$, 
the chiral critical line, but only the Roberge-Weiss critical lines~\footnote{
Error bars for the determinations at $\beta = 5.10$ and 
on the critical lines ($\theta = \pi/3$ and $\theta = \pi$) 
are probably underestimated.}.

We have also performed runs at fixed $\mu_I$ and variable $\beta$ to 
look for other locations of the chiral line in the $T$--$\mu_I$ plane,
a detailed summary of all results will be presented elsewhere~\cite{else}.
We present, in Fig.~\ref{diagfig}, a sketch of the phase diagram
in the $\beta$--$\mu_I$ plane, as emerges from our data
and by exploiting the above mentioned symmetries.
We can distinguish a region where chiral symmetry is spontaneously 
broken (indicated as IV in the figure) and three regions (I, II and III),
which correspond to different $Z_3$ sectors and repeat periodically, 
where chiral symmetry is restored. The chiral critical line
separates region IV from other regions, while the Roberge-Weiss critical lines
separate regions I,II and III among themselves.
In QCD with 4 staggered flavours, $a m = 0.05$ and $\mu_I = 0$ , 
the phase transition is expected
to be first order~\cite{brown}: assuming it continues to be first
order also at $\mu_I \neq 0$, we expect all the regions to be separeted by
first order critical lines.

\begin{figure}[!t]
\vspace{0cm}
\epsfxsize=27pc 
\epsfbox{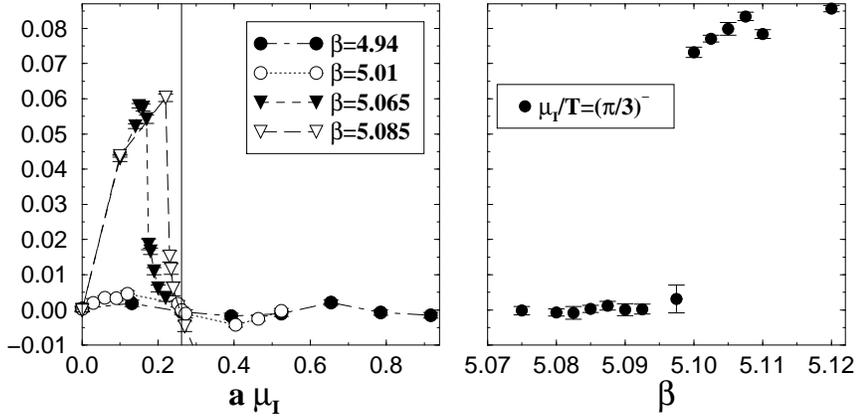}
\vspace{0.20cm}
\caption{Imaginary part of the barion density as a function of
$\mu_I$ for different values of $\beta$ (left--hand side), and 
as a function of $\beta$ at $\theta = \mu_I/T =  \frac{\pi^-}{3}$
(right-hand side).}
\label{barfig}
\vspace{-0.6cm}
\end{figure}

It is interesting to illustrate the determination of the endpoint
of the Roberg-Weiss critical line, $\beta_E = 5.097(2)$. We have performed
a simulation at exactly $\theta = \pi/3$: starting thermalization at an high
value of $\beta$ from a zero field configuration we always stay 
on the border of region I, since on the $16^3 \times 4$ lattice that we 
have used it is practically impossible to flip into region II through
the critical line. As we decrease $\beta$, always staying on the 
border of region I, we will meet the chiral critical
line at $\beta = \beta_E$. Various quantities can obiouvsly be considered 
to signal this transition, but it is interesting to notice that in this case
the baryon density can be taken as an exact order parameter.
Indeed the baryon density, 
$\langle b \rangle = \frac{T}{V} \frac{\partial}{\partial \mu} \ln Z$,
is an odd function of $\mu$, being $Z$ an even function. Therefore, 
for an imaginary chemical potential, $\langle b \rangle$
is also purely imaginary and an odd function of $\mu_I$. This, combined
with the periodicity in $\mu_I$, leads to the expectation that
$\langle b \rangle (\theta = \frac{\pi^-}{3}) = - \langle b \rangle 
(\theta = \frac{\pi^+}{3})$.
The last relation clearly implies that $\langle b \rangle = 0$ at 
$\theta = \pi/3$, unless $\langle b \rangle$ is not continuous on that
point. Thus a non-zero value of $\langle b \rangle = 0$ at 
$\theta = \frac{\pi^-}{3}$ implies the presence of the Roberge-Weiss
critical line. 
On the right hand side of Fig.~\ref{barfig} the imaginary part of
$\langle b \rangle$ at $\theta = \frac{\pi^-}{3}$ is plotted as a function
of $\beta$: one can clearly see a transition from a zero to a non-zero
expectation value, which permits the determination of $\beta_E$.
We have verified that the transition through $\beta_E$ is also
visible in the chiral condensate: this implies that the Roberge-Weiss
critical line ends on the chiral critical line. 
On the left hand side of Fig.~\ref{barfig} we present instead 
the imaginary part of $\langle b \rangle$ as a function of
$\mu_I$ for different values of $\beta < \beta_E$: in this case
$\langle b \rangle$ is always zero and continuous 
at $\theta = \frac{\pi}{3}$, but
it is interesting to note how it starts developing the discontinuity
as $\beta \to \beta_E$.

In order to translate results for $\beta_c(\mu_I)$
into results for the physical critical temperature, $T_c(\mu_I)$, 
we need the lattice spacing, $a = a(\beta)$, in physical units. 
For instance using the values $a(5.04) = 0.30(2)$ fm  and 
$a(5.097) = 0.272(10)$~fm~\cite{topol} we obain
$T_c = 164(10)$MeV and $T_E = 181(7)$MeV. 

Finally we notice that for high temperatures and
away from the critical lines, physical quantities show a clear
non-zero dependence on the imaginary chemical potential, which 
is encouraging in starting the program of fitting the first terms 
of their Taylor expansion in $\mu_I$ and performing the analytic continuation
to real chemical potential.
As an example we have reported, in Fig.~\ref{psifig}, the result
of a quadratic fit in $\mu_I$ for the chiral condensate at $\beta = 5.10$.

\section{Summary and discussion}

We have clarified the phase structure in the imaginary chemical
potential -- temperature plane for full QCD with four staggered flavours.
In particular we have confirmed the existence of the Roberge Weiss 
critical lines, located their endpoints and assessed 
their interplay with the chiral critical lines.

We have checked that data at high temperature and small imaginary
chemical potential can be safely fitted by a polinomial, when away
from phase transition lines, making the analytic continuation to real $\mu$ feasible.

The physical interesting region of high temperature and small 
chemical potential can now be studied by essentially three different 
techniques: 1) Direct calculations of derivatives;  2) Reweighting;  
3) Analytic continuation from imaginary $\mu$. 
Each method has its own merits and
limitations, and cross checks among the three approaches should
produce reliable results.

\section*{Acknowledgments}
This work has been partially supported by MIUR.
We thank the computer center of ENEA for providing us with time on
their QUADRICS machines.

\end{document}